\documentclass[aps,prb,a4paper,preprint,showpacs]{revtex4}
\usepackage{amsmath}
\usepackage{amsfonts}
\usepackage{amssymb}
\usepackage{slashbox}
\usepackage{graphicx}
\usepackage{amsbsy}

\begin{document}
\title{Spin-dependent electron grating effect from helical magnetization
in multiferroic tunnel junctions }
\author{Rui Zhu\renewcommand{\thefootnote}{*}\footnote{Corresponding author. Electronic address:
rzhu@scut.edu.cn}}
\address{Department of Physics, South China University of Technology,
Guangzhou 510641, People's Republic of China }

\begin{abstract}

In multiferroic oxides with a transverse helical magnetic order, the
magnetization exchange coupling is sinusoidally space-dependent. We
theoretically investigate the spin-dependent electron grating effect
in normal-metal/helical-multiferroic/ferromagnettic heterojunctions.
The spin wave vector of the spiral can be added or subtracted from
the electron spacial wave vector inducing spin-conserved and
spin-flipped diffracted transmission and reflection. The predicted
grating effect can be controlled by magnetization exchange coupling
strength, the helicity spatial period, and the magnetization of the
ferromagnetic layer.

\end{abstract}


\maketitle

\narrowtext

\section{Introduction}

In optics, a diffraction grating is an optical component with a
periodic structure, which splits and diffracts light into several
beams traveling in different directions. The directions of these
beams depend on the spacing of the grating and the wavelength of the
light\cite{Ref1}. Similar effect can be found in electron transport
as an effect of particle-wave duality. Electron diffraction is most
frequently used in high-energy transmission electron microscope to
study the crystal structure of solids. In mesoscopic tunnel
junctions, the electron energy is determined by the Fermi energy of
the reservoirs, which could be in the order of meV. In this energy
scale, the electron wave length can be as large as several
nanometers. Transport diffraction can be observed in nano-scale
gratings. Recently, optical transient spin-grating spectroscopy was
used to measure the lifetime of spin polarization waves in
spin-orbit-coupled semiconductor quantum wells\cite{Ref2}.
Diffraction by transient helical spin wave generated by exciting a
2DEG with two non-colinear beams from a femtosecond laser can be
detected by a probe pulse\cite{Ref2}. Preceding experimental
realization, spin propagation theories in semiconductor 2DEG with
Rashba spin-orbital coupling are intensively discussed\cite{Ref3,
Ref4}.

The coexistence of coupled electric and magnetic order parameters in
multiferroics\cite{Ref5} (MF) holds the promise of new opportunities
for device fabrications\cite{Ref6}. Our interest is focused on the
helimagnetic MF, in which the magnetization exchange coupling is
sinusoidally space-dependent. In real space, the eigen spinor of the
helimagnet is a static spin wave pointing along the helical
magnetization. Using a space-dependent gauge transformation, it is
shown that the topology of the local helical magnetic moments in
these materials induces a resonant, momentum-dependent spin-orbit
interaction (SOI)\cite{Ref7}. Jia \emph{et al.} found that the
momentum dependence of the SOI is analogous to semiconductor-based
2DEG with the Rashba and Dresselhaus spin-orbit interaction being
equal\cite{Ref7}. When the strength of the two dominant SOI are
equal, SU(2) symmetry is restored, giving rise to a persistent
helical spin density wave confirmed by transient spin-grating
spectroscopy\cite{Ref2}. This analogy suggests the existence of
similar effect in MF helimagnets. Recently, electron spin resonance
in chiral helimagnet was proposed\cite{Ref8}. In this work, we
investigate the spin-resolved transmission properties in
normal-metal (NM)/helical-MF/ferromagnetic (FM) heterojunctions.
Spin-flipped grating effect is found in electron transmission as a
result of the sinusoidally-space-dependent magnetization in the MF
layer.

\section{Theoretical formulation}

The MF tunnel junction we consider follows that proposed by Jia et
al.\cite{Ref6}. It consists of an ultrathin helical-MF barrier
sandwiched between a NM layer and a FM conductor. The ferroelectric
polarization ${\bf{P}}$ in the MF barrier creates, in general,
surface charge densities $ \pm \left| {\bf{P}} \right|$ which are
screened by the induced charge at the two metal
electrodes\cite{Ref9, Ref10}. Potential drop in the ferroelectric
phase of ${\rm{TbMnO}}_{\rm{3}} $ generated by the depolarizing
field in the thin-film MF barrier is estimated to be on the energy
scale of millielectron volt, which is much smaller than any other
relevant energy scale in the system\cite{Ref6, Ref11}. In the
present study, we neglect this potential modification and assume
that the barrier potential has a rectangular shape with the height
$V_{0}$. Based on this approximation, the Hamiltonians governing the
carrier dynamics in the two electrodes and the oxide insulator have
the following form:
\begin{equation}
\begin{array}{c}
 \begin{array}{*{20}c}
   {H_{NM}  =  - \frac{{\hbar ^2 }}{{2m_e }}\nabla ^2 ,} & {z < 0,}  \\
\end{array} \\
 \begin{array}{*{20}c}
   {H_{MF}  =  - \frac{{\hbar ^2 }}{{2m^* }}\nabla ^2  + J{\bf{n}}_{\bf{r}}  \cdot {\bf{\sigma }} + V_0 ,} & {0 \le z \le d,}  \\
\end{array} \\
 \begin{array}{*{20}c}
   {H_{FM}  =  - \frac{{\hbar ^2 }}{{2m_e }}\nabla ^2  - \Delta {\bf{m}} \cdot {\bf{\sigma }},} & {z > d,}  \\
\end{array} \\
 \end{array}
 \end{equation}
where $V_{0}$ and $d$ are the height and width of the potential
barrier, and $m_{e}$ is the free-electron mass. $m^{*}$ is the
effective electron mass of the oxide (${{m^* } \mathord{\left/
 {\vphantom {{m^* } {m_e }}} \right.
 \kern-\nulldelimiterspace} {m_e }} \approx 10$) and ${\bf{\sigma }}$
is the Pauli vector. ${\bf{m}} = \left[ {\sin \theta \cos \phi
,\sin\theta \sin \phi ,\cos \theta } \right]$ is a unit vector
defining the magnetization direction in the FM with respect to the
[100] crystallographic direction. $\Delta$ is the half width of the
Zeeman splitting in the FM electrode. $J{\bf{n}}_{\bf{r}} $ is the
exchange field, where ${\bf{n}}_{\bf{r}} $ is given by the MF oxide
local magnetization at each spiral layer (labeled by an integer $l$)
along the $z$ axis\cite{Ref12}, i.e., ${\bf{n}}_{\bf{r}}  = \left( {
- 1} \right)^l \left[ {\sin \theta _r ,0,\cos \theta _r } \right]$
with $\theta _r = \bar q_m  \cdot {\bf{r}}$ and $\bar q_m  = \left[
{\bar q,0,0} \right]$ being the spiral spin-wave vector.
The
physical picture behind the term $H_{MF}$ in Eq. (1) is that a
tunneling electron experiences an exchange coupling at the sites of
the localized noncollinear magnetic moments within the barrier. In
effect this acts on the electron as a nonhomogenous magnetic field.

Experimental observations\cite{Ref13, Ref14, Ref15} indicate that
thin-film MF can retain both magnetic and ferroelectric properties
down to a thickness of 2 nm (or even less). We consider ultrathin
tunneling barriers that can be approximated by a Dirac-delta
function\cite{Ref16}. The helical MF barrier reduces to a plane
barrier. Hamiltonian in the MF layer can be rewritten as
\begin{equation}
H_{MF}  =  - \frac{{\hbar ^2 }}{{2m^* }}\nabla ^2  + \left( {\tilde
J{\bf{n}}_r  \cdot {\bf{\sigma }} + V_0 d} \right)\delta \left( z
\right),
\end{equation}
where we assume a single spiral layer with $l=0$. $\tilde J =
\left\langle {J\left( z \right)} \right\rangle d $ refers to space
and momentum averages with respect to the unperturbed states at the
Fermi energy. It should be noted that the helimagnetic exchange
coupling is sinusoidally space dependent. A multichannel-tunneling
picture should be considered and integer numbers of the helical wave
vector $\bar q$ could be absorbed or emitted in transmission and
reflection. With a plane wave incidence, the general wave function
of the incident, transmitted and reflected electrons can be written
as
\begin{equation}
\psi _{NM}^\sigma  \left( x,y,z \right) = e^{ik_x x} e^{ik_y y}
e^{ik_z z} \chi _\sigma   + e^{ik_y y} \sum\limits_{\sigma ',n}
{r_n^{\sigma \sigma '} e^{i {k_{x}^{n}  } x} e^{ - ik_z^n z} \chi
_{\sigma '} } ,
\end{equation}
\begin{equation}
\psi _{FM}^\sigma  \left( x,y,z \right) = e^{ik_y y}
\sum\limits_{\sigma ',n} {t_n^{\sigma \sigma '} e^{i {k_x^n} x}
e^{ik_z^{n\sigma '} z} \chi _{\sigma '} } ,
\end{equation}
with $k_z  = \sqrt {2mE - \hbar ^2 k_y^2 } \cos
\theta_{xz} / \hbar ,$ $k_x  = \sqrt {2mE - \hbar ^2 k_y^2 } \sin
\theta_{xz}/ \hbar ,$
and $\theta_{xz}$ the incident angle in the $x$-$z$ plane. Here,
$n$ is an integer ranging from $ - \infty $ to $  \infty $ indexing
the diffraction order. And $
k_z^n  = \sqrt {2mE - \hbar ^2 k_y^2  - \hbar ^2 (k_x^n) ^2 }/\hbar $, $k_z^{n\sigma }  = \sqrt {2mE + \sigma \Delta  - \hbar ^2
k_y^2  - \hbar ^2 (k_x^n) ^2 }/ \hbar $, $k_x^n  = k_x  + n\bar q$.
The spinors are introduced as $\chi _ +   = \left(
{\begin{array}{*{20}c}
   1 , &
   {tg\frac{\theta }{2}e^{i\phi } }
\end{array}} \right)^{\rm T} $ and ${\chi _ -   = \left( {\begin{array}{*{20}c}
   1,  &
   { - ctg\frac{\theta }{2}e^{i\phi } }
\end{array}} \right)^{\rm T} }$,
corresponding to an electron spin parallel ($\sigma =+$) and
antiparallel ($\sigma =-$) to the magnetization direction in the FM
electrode. In the plane ($x$-$y$ plane) perpendicular to the
transport direction ($z$-axis), free motion of the electron is
assumed. Therefore, only propagating modes in the $x$-direction is
considered. And $k_y$ is conserved under translational invariance.

The reflection ($r_n^{\sigma \sigma '}$) and transmission
($t_n^{\sigma \sigma '}$) amplitude in the $n$th diffraction order
can be numerically obtained from the continuity conditions\cite{Ref6, Ref16} for $\Psi
(x,y,z)$ at $z=0$.
\begin{equation}
\Psi _{NM}^\sigma  \left( {x,y,0^ -  } \right) = \Psi _{FM}^\sigma
\left( {x,y,0^ +  } \right),
\end{equation}
\begin{equation}
\begin{array}{c}
 \frac{{\hbar ^2 }}{{2m_e }}\left. {\frac{{\partial \Psi _{NM}^\sigma  \left( {x,y,z} \right)}}{{\partial z}}} \right|_{z = 0^ -  }  + \left[ {V_0 d + {\bf{\tilde w}}\left( {\theta _r } \right)} \right]\Psi _{NM}^\sigma  \left( {x,y,0^ -  } \right) \\
  = \frac{{\hbar ^2 }}{{2m_e }}\left. {\frac{{\partial \Psi _{FM}^\sigma  \left( {x,y,z} \right)}}{{\partial z}}} \right|_{z = 0^ +  } , \\
 \end{array}
\end{equation}
with
\[
{\bf{\tilde w}}\left( {\theta _r } \right) = \tilde J\left[
{\begin{array}{*{20}c}
   {\cos \theta _r } & {\sin \theta _r }  \\
   {\sin \theta _r } & { - \cos \theta _r }  \\
\end{array}} \right].
\]
The continuity equation can be expressed in the equations of each
diffracted order. Transmissivity of a spin-$\sigma$ electron through the MF tunnel
junction with the incident wave vector $[k_x,k_y,k_z]$ to the $n$-th
diffracted order and spin-$\sigma '$ channel with the outgoing wave
vector $[k_x^n,ky,k_z^{n\sigma '}]$ reads
\begin{equation}
T_n^{\sigma \sigma '} \left( {E,k_y ,\theta _{xz} } \right) =
\frac{{ \left| {k_z^{n\sigma '} } \right|  }}{{ k_z }}\left|
{t_n^{\sigma \sigma '} } \right|^2.
\end{equation}

\section{Numerical results and interpretations}

We consider the diffracted transmission properties of the NM/helical
MF/FM junctions. In numerical calculations, the NM Fermi energy
$E_F$ is chosen\cite{Ref6} to be $5.5$ eV. The spatial average of
the helimagnetic exchange coupling strength $\tilde J = 0.2$
${\rm{eV}} \cdot {\rm{nm}}$, which is reasonable compared to the
Fermi energy. Period of short-period helimagnets\cite{Ref17} is $3$-$6$ nm and
of long-period helimagnets is $18$-$90$ nm. In our model
we consider the period to be $10$ nm and hence $\bar q=2 \pi /10$
$\texttt{nm} ^{-1}$. The magnetization direction in the FM is fixed
to be $\theta =1$ and $\phi =0.5$ in radian. Zeeman splitting in the
FM electrode $\Delta =2$ eV. Barrier height of the MF oxide plane
$V_0 =0.5$ eV and width $d=2$ nm. We consider 5 diffraction orders,
which is sufficient as higher orders decrease exponentially.

During transmission, the incident electron with wave
vector $[k_x,k_z]$ would absorb or emit $n \bar q$ from the
helimagnet and be diffracted into tunnels with wave vector
$[k_x^n,k_z^{n \sigma}]$. The diffraction level can be defined by
the number $n$ being sequel positive and negative integers following
the definition in optical gratings. $n=0$ labels the
zero-order transmission channel with the transmission direction the
same to the incidence. $n=\pm 1$ labels the $\pm 1$-order
transmission channel with the transmission direction shift an angle
to $[k_x \pm \bar q,k_z^{\pm 1 \sigma }]$, and etc. The diffraction
tunnel number correspondence is illuminated in Fig. 1. The incident
wave vector is set to be $k_x=\bar q$ to make figure scale brief and
the diffraction phenomenon is analogous for all incident $k_x$.

Numerical results of the transmission diffraction effect are shown
in Fig. 2. The incident wave vector is set to be $k_x = \bar q$ and
analogous diffraction effect can be found for continuously all $k_x$. From the
transmissivity $T^{\sigma \sigma '}$ [$T_n^{\sigma \sigma '}$ in Eq.
(7)] plotted in Fig. 2, it can be seen that the spin-conserved
transmission is a direct grating effect with the transmission
amplitude exponentially decreasing as the diffraction order is
increased (see also insets of Fig. 2). The transmission of one-way
incident light through sinusoidal gratings is delta-function-like
strict lines. Analogously, direction of transmission of one-way
incident electron through sinusoidal helimagnet is discrete strict
lines of different grating orders and the spin is conserved or
flipped. The spin-flipped transmission is an effect of the
electron-helimagnet coupling. As a result, $\pm 1$ order magnifies
in spin-flipped transmissivity. For arbitrary $\textbf{m}$ relative
to the chirality of the helimagnet, $+ 1$ and $-1$ order
transmission may not be symmetric. Physically, an electron with spin
polarization along the FM magnetization is transmitted in different
direction with its spin rotating an angle as the cartoon Fig. 3
shows.

Results of an arbitrary FM magnetization is shown in Fig. 2. We also
considered some special FM configurations. Numerically calculation
demonstrates that for $\textbf{m}$ pointing to the $z$-direction
($\theta =0$), an incident spin-down electron would be completely
transmitted to $\pm 1$ order with all the spin flipped and
transmission in the other channels is several orders smaller.
Reversely, for $\textbf{m}$ pointing to the $-z$-direction ($\theta
= \pi $), an incident spin-up electron would be completely
transmitted to $\pm 1$ order with all the spin flipped and
transmission in the other channels is several orders smaller.
Opposite channel priority can also be found when the chirality of
the helimagnet is turned over. Along this physical track, for $\textbf{m}$ lying in the $x$-direction ($\theta =\pi
/2$, $\phi =0$), $\pm 1$ order transmission magnifies in
spin-flipped channels with $+ 1$ and $-1$ orders symmetric. For
$\textbf{m}$ lying in the $y$-direction ($\theta =\pi /2$, $\phi
=\pi /2$), spin-conserved grating is extremely suppressed and
spin-flipped transmission would only occur in $+ 1$ order for
spin-up incidence and in $- 1$ order for spin-down incidence. These
results can be interpreted by the effect of the exchange coupling in
the helimagnet. When the spin polarization is along the helimagnet
spiral axis ($z$ direction in our structure), the grating "seen" by
the incident spin-up electron and that "seen" by the spin-down
electron is in opposite modulation, hence spin-flipped grating
effect is enhanced. When the spin polarization is perpendicular to
the helimagnet spiral plane ($x$-$z$ plane in our structure) spin
grating effect is extremely suppressed.

In the NM/helical-MF/FM junction setup, all adjustable parameters
can induce difference phenomenons in the electron diffracted
transmission. In above discussions a short-period helimagnet (with
period $10$ nm) is considered. For long-period helimagnet ($18$-$90$
nm), grating effect would be extremely diminished. Above we
considered a small exchange coupling strength in the helical MF
oxide. For strong coupling strength comparable to the Fermi energy,
spin grating effect would dominate direct transmission making the
diffraction pattern more prominent.

\section{Conclusions}

We theoretically investigate the spin-dependent electron grating
effect induced by scattering from a sinusoidal helimagnet thin film
sandwiched between NM and FM electrodes. An incident electron with
spin polarization parallel or antiparallel to the FM magnetization
is diffracted into different direction with its spin rotating an
angle resembling an optical grating effect. The diffraction
phenomenon can be tuned by external parameters such as the
magnetization exchange coupling strength, the helicity spatial
period, and the magnetization of the ferromagnetic layer.

\section{Acknowledgements}

The author acknowledges enlightening discussions with Jamal Berakdar, Wen-Ji Deng, and Zhi-Lin Hou.
This project was supported by the National Natural Science
Foundation of China (No. 11004063) and the Fundamental Research
Funds for the Central Universities, SCUT (No. 2012ZZ0076).

\clearpage

\begin{figure}[h]
\includegraphics[height=10cm, width=8cm]{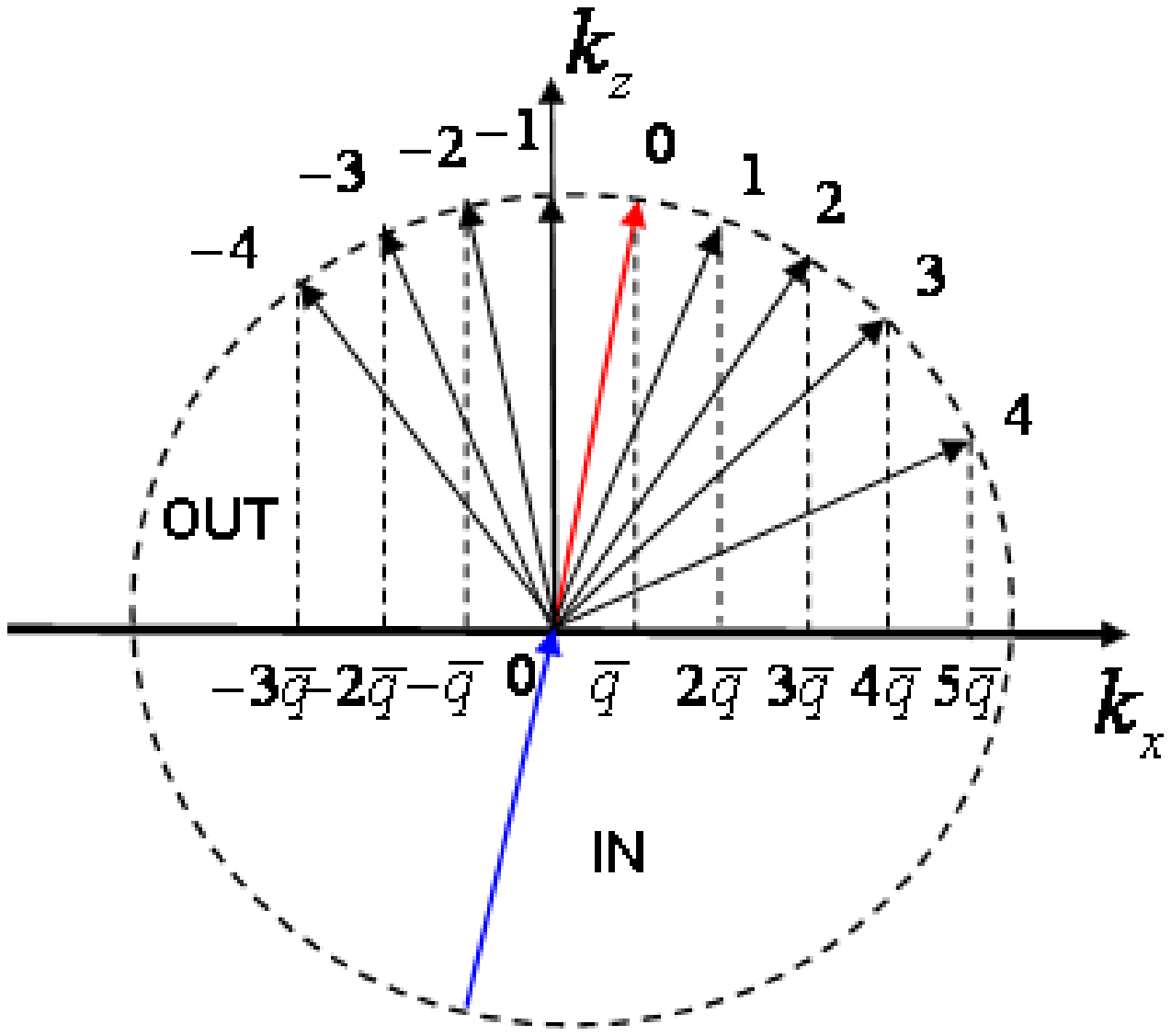}
\caption{Different tunnels in the wave vector space. The incident
beam travels with wave vector $(k_{x}=\bar q,k_{z})$. The
transmitted beam traveling in the same direction of the incident
beam is indexed as tunnel 0. The transmitted beam with wave vector
$(k_x^n=k_x+n \bar q,k_z^n)$ are indexed as tunnel $n$ with $n$
being sequel positive and negative integers following the definition
in optical grating effect.}
\end{figure}

\clearpage

\begin{figure}[h]
\includegraphics[height=8cm, width=10cm]{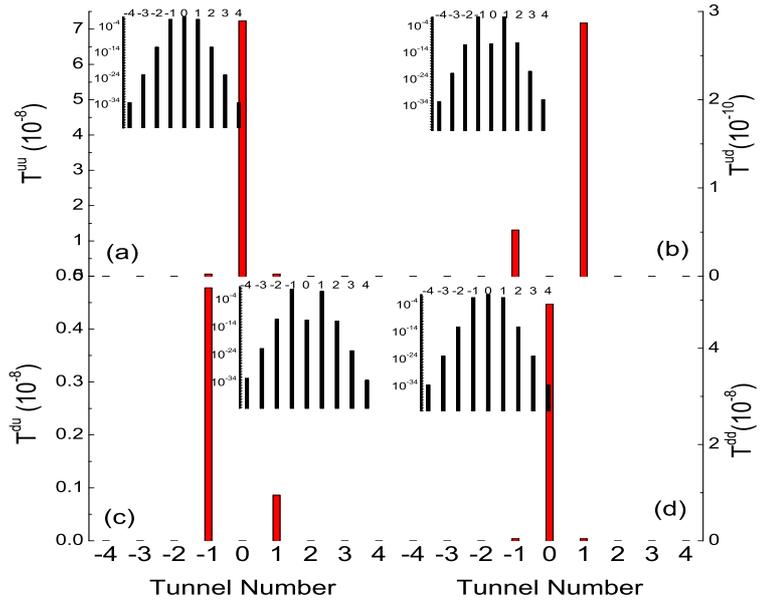}
\caption{Spin-conserved and spin-flipped transmission in different
tunnels $(k_x^n=k_x+n \bar q,k_z^n)$ of an incident spin-up electron
with wave vector $(k_{x}=\bar q,k_{z})$. Numbers label the tunnel
indexes $n$. Four panels indicate transmission $T^{ \uparrow
\uparrow } $ ($T^{uu}$), $T^{ \uparrow \downarrow } $ ($T^{ud}$),
$T^{ \downarrow \uparrow } $ ($T^{du}$), and $T^{ \downarrow
\downarrow } $ ($T^{dd}$), respectively. Insets are logarithms of
the transmission to show the slight effect of high diffraction
orders. The used numerical values are $E_{F}=5.5$ eV, $V_{0}=0.5$
eV, $d=2$ nm, $\bar q=2 \pi /10$ $\texttt{nm}
 ^{-1}$, $\phi =0.5 $ radian, $\theta =1$ radian, $k_{x}= \bar q$, $k_{y}=2 \bar q $,
 and $\tilde J =
0.2$ ${\rm{eV}} \cdot {\rm{nm}}$.}
\end{figure}

\clearpage

\begin{figure}[h]
\includegraphics[height=14cm, width=10cm]{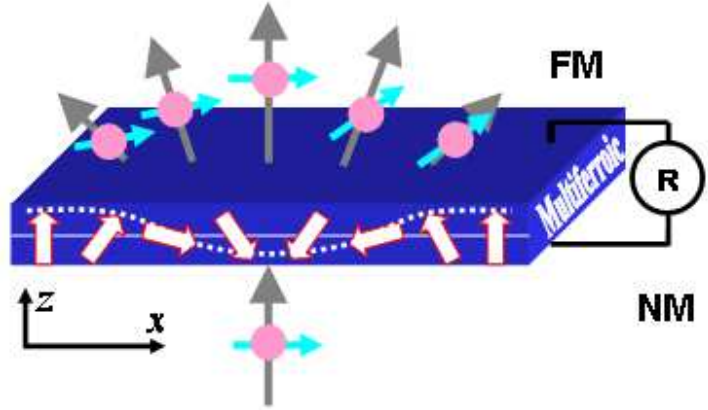}
\caption{A cartoon sketching a spin-dependent electron grating
effect from helical magnetization in multiferroic tunnel junctions.
The helical spin structure in a helimagnet diffracts a
spin-polarized electron beam into several beams traveling in
different directions with all beams containing spin-flipped
components. The green arrows indicate the transmitted spin
polarization. Helicity of the multiferroic oxide is in white
arrows.}
\end{figure}

\clearpage

\end{document}